\documentclass[conference]{IEEEtran}

\usepackage{fst}
\usepackage{cite}

\begin{document}

\title{Ultra-Sparse Non-Binary LDPC Codes\\ for Probabilistic Amplitude Shaping}
\author{%
\IEEEauthorblockN{Fabian Steiner\IEEEauthorrefmark{1}, Gianluigi Liva\IEEEauthorrefmark{2}, Georg B\"ocherer\IEEEauthorrefmark{1}}
\IEEEauthorblockA{\IEEEauthorrefmark{1}Institute for Communications Engineering\\Technical University of Munich\\Email: \{fabian.steiner, georg.boecherer\}@tum.de}
\IEEEauthorblockA{\IEEEauthorrefmark{2}Deutsches Zentrum f\"ur Luft- und Raumfahrt\\ We\ss{}ling, Germany\\Email: gianluigi.liva@dlr.de}}

\tikzset{%
    funblock/.style = {draw,top color = gray!1,bottom color = gray!10,rounded corners,rectangle,text centered,minimum height = 20, minimum width = 25,inner sep=3}
}

\markboth{}{}%


\maketitle

\begin{abstract}
This work shows how non-binary low-density parity-check codes over $\F_{2^p}$
can be combined with probabilistic amplitude shaping (PAS)~(B\"ocherer, \textit{et al.}, 2015), which 
combines forward-error correction with non-uniform signaling for power-efficient communication.
Ultra-sparse low-density parity-check codes over $\F_{64}$ and $\F_{256}$ gain
\SI{0.6}{dB} in power efficiency over state-of-the-art binary LDPC codes
at a spectral efficiency of \SI{1.5}{bits per channel use} and a blocklength of \num{576} bits. The simulation results are
compared to finite length coding bounds and complemented by density evolution analysis.
\end{abstract}

\section{Introduction}
\label{sec:intro}

Bandwidth-limited communication systems must use higher-order modulations such as
\ac{QAM} or \ac{APSK} to provide the required \ac{SE} for high data rates. In most standards, these modulation
formats are combined with uniform signaling, i.e., all points in the constellation are transmitted with the same probability.
This approach loses up to \SI{1.53}{dB}~
\cite[Sec.~IV]{forney_efficient_1984} in power efficiency and lacks flexibility because it restricts the choice of \acp{SE} to supported modulation order
and code rate combinations (modcods). Recently, \ac{PAS} was proposed \cite{bocherer_bandwidth_2015} to overcome these 
shortcomings: \ac{PAS} closes the shaping gap and allows seamless rate adaptation by adjusting the channel input distribution.
The enabling device is the \ac{DM}~\cite{schulte_constant_2016} which transforms uniform input bits
into arbitrarily shaped sequences. The practicality of this scheme has been demonstrated in a number of optical field trials~\cite{buchali_rate_2016,
idler_field_2017,cho_trans-atlantic_2017} and has been proposed to 3GPP for inclusion in the upcoming 5G standard~\cite{huawei_tdoc}. In all
these works, either \emph{binary} \ac{LDPC}, turbo or polar codes~\cite{prinz_polar_2017} have been employed.

One important scenario in the upcoming standards is \emph{ultra reliable low-latency communication} (uRLLC)~\cite{durisi_toward_2016}, which 
requires \ac{FEC} with small blocklengths and very low bit error rates to guarantee the required reliability. Non-binary codes are particularly 
good codes in this regime of operation~\cite{chang_non-binary_2012,liva_short_2013,dolecek_non-binary_2014}.

The combination of \ac{PAS} and non-binary codes was suggested in \cite{boutros_probabilistic_2017}. 
Herein, the authors propose a new design for circular \ac{QAM} constellations that can be used with non-binary codes over
prime fields of order larger than two. 

In this work, we propose a different strategy and consider non-binary codes over the extension field $\F_q$ with $q = 2^p$. We illustrate
the principle by ultra-sparse \ac{NB-LDPC}~\cite{davey_low-density_1998,hu_cycle_2003} codes, which have shown an excellent
performance on the \ac{biAWGN} channel for short blocklengths~\cite{poulliat_design_2008}. These codes are also known as cycle codes~\cite{peterson_error-correcting_1972} 
and have constant variable and check node degrees, where the former is fixed to two. Codes over $\F_{2^p}$ also allow for 
low complexity decoding using the \ac{HT}~\cite{barnault_fast_2003}.

This paper is structured as follows. Sec.~\ref{sec:prelim} reviews the system model and introduces achievable rate expressions. In Sec.~\ref{sec:nb_with_pas},
we show how \ac{NB-LDPC} codes can be combined with PAS. Sec.~\ref{sec:numerical_results} provides numerical simulation results and a
comparison with binary \ac{LDPC} codes. We conclude in Sec.~\ref{sec:conclusion}.

\section{Preliminaries}
\label{sec:prelim}

\subsection{System Model}
\label{sec:system_model}

Consider transmission over a real-valued \ac{AWGN} channel
\begin{equation}
 Y_i = X_i + Z_i\label{eq:system_model}
\end{equation}
for $i=1,\ldots,n$ channel uses. The realizations for the channel input $X_i$
are taken from a scaled $M=2^m$-ary \ac{ASK} constellation $\cX = \{\pm1,\pm3,\ldots,\pm (M-1)\}$ such that $\E{X_i^2} = 1$. The results extend directly 
to \ac{QAM}, where we use \ac{ASK} for the in-phase and quadrature transmission. 
The noise $Z_i$ is a Gaussian random variable with zero mean and variance $\sigma^2$. The \ac{SNR} is $1/\sigma^2$.
As the channel is memoryless, we drop the index $i$ and denote the governing
channel law as $p_{Y|X}$. 
The mutual information maximizing distribution under an average power constraint is a zero mean Gaussian input $X$ with unit variance,
and it yields the capacity expression
\begin{equation}
 \sfC_\tawgn(\text{SNR}) = \frac{1}{2}\log_2(1+\SNR)\label{eq:cap_awgn}.
\end{equation}
To approach $\sfC_\tawgn$ with discrete signaling, either \ac{GS} or \ac{PS} can be employed
to mimic a Gaussian shape. In \cite{Stei1702:Comparison}, we demonstrated the superiority of \ac{PS}
for practical coded modulation scenarios, and we use \ac{PAS} therefore. Numerical 
comparisons in \cite[Table~I]{bocherer_bandwidth_2015} show that \ac{MB} distributions~\cite{kschischang_pasupathy_maxwell}
\begin{equation}
 P_X(x) \propto \exp(-\nu x^2)
\end{equation}
are nearly optimal. They are also natural choices for power efficient communication, as they
are the solution to the problem of minimizing the average power of the channel inputs subject to an entropy constraint. 

An achievable rate for \ac{SMD} is
\begin{equation}
 R_\tsmd(\SNR; P_X) = \I(X;Y)\label{eq:rsmd}
\end{equation}
where $\I(X;Y)$ is the mutual information between the channel input $X$ and channel output $Y$.
To use binary codes, we label each constellation point $x\in\cX$ with an $m$-bit binary label, 
i.e., $\beta : \cX \to \{0,1\}^m$ and $\beta(x) = B_1B_2\ldots B_m = \vB$. A \ac{BRGC}~\cite{gray1953pulse} usually 
performs well. An achievable rate for \ac{BMD} is given by~\cite{bocherer2014achievable}
\begin{equation}
 R_\tbmd(\SNR; P_X) = \left[\entr(\vB) - \sum_{i=1}^m \entr(B_i|Y)\right]^+\label{eq:rbmd}.
\end{equation}
We denote the Shannon limits for \eqref{eq:cap_awgn}, \eqref{eq:rsmd} and \eqref{eq:rbmd}
for a fixed \ac{SE} as $\SNR_{\text{CAP}}$, $\SNR_{\tsmd}$ and $\SNR_{\tbmd}$, i.e.,
$R_\tsmd(\SNR_\tsmd; P_X) = R_\tbmd(\SNR_\tbmd; P_X) = C_\tawgn(\SNR_\text{CAP})$.

\subsection{Distribution Matching}
\label{sec:dm}

A \ac{DM}~\cite{schulte_constant_2016} transforms uniformly distributed bits into shaped symbols.
In our setup, we employ a fixed-to-fixed length \ac{DM} which maps $k$ input bits to $n$ output
symbols from the amplitude set $\cA$ of the $M$-ASK constellation. The mapping is invertible, so the input can 
be recovered from the output. Both the desired distribution $P_A$ and the output length $n$ serve as an input parameter to the \ac{DM}.
The matcher rate is 
\begin{equation*}
 R_\tdm = k/n\quad \left[\frac{\text{bits}}{\text{output symbols}}\right].
\end{equation*}

\begin{figure*}
 \footnotesize
 \centering
 \includegraphics{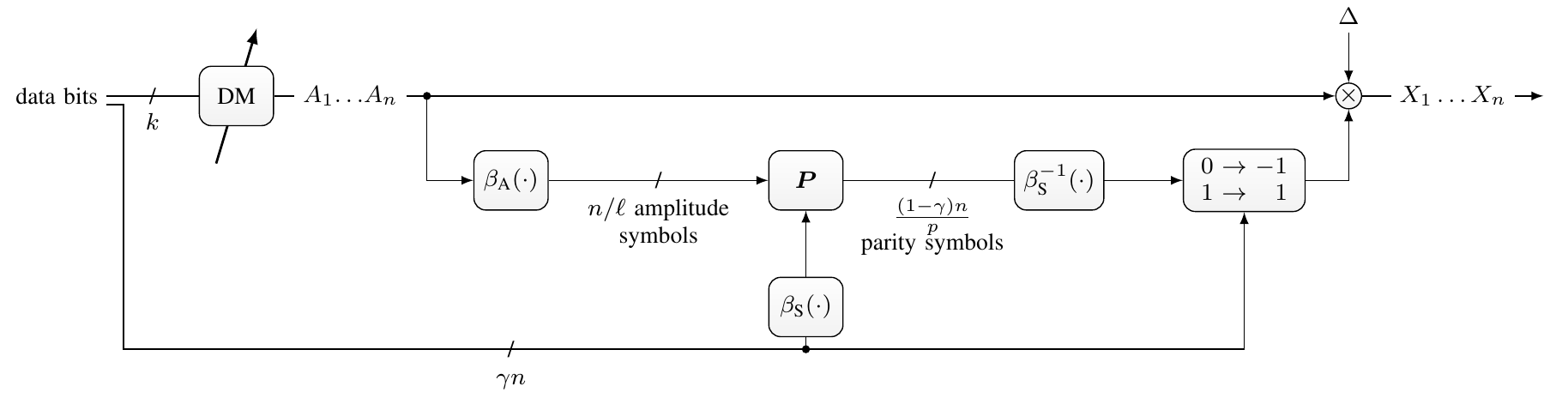}
 \caption{System model of PAS for NB $\F_{q}$ codes and $q = 2^p$.}
 \label{fig:pas}
\end{figure*}

\section{Non-Binary LDPC Codes over $\F_q$ with PAS}
\label{sec:nb_with_pas}

An \ac{NB-LDPC} code $\cC$ is defined as the nullspace of the sparse parity-check matrix $\vH$ of dimension $m_\tc\times n_\tc$ 
where the non-zero entries $h_{ji}$ of $\vH$ are taken from the finite field $\setF_{q}$, i.e.,
$\cC = \left\{\vc\in\setF_{q}^{n_\tc}: \vc\vH^{\tT} = \zeros\right\}$. The parity-check matrix $\vH$ can be represented by
a bipartite graph, called the Tanner graph~\cite{tanner_recursive_1981}. Every codeword symbol $c_i$ is represented by one of the $n_\tc$ variable nodes $V_i$ in
the graph. The $m_\tc$ linear constraints are represented by check nodes $C_j$. If the edge label $h_{ji}$ is non-zero, then there is an edge between $V_i$ and $C_j$.
Variable and check node $V_i$ and $C_j$ have degree $d_{\tv,i}$ and $d_{\tc,j}$, respectively, where the degree specifies
the number of associated edges. In the following, we use a special class of \ac{NB-LDPC} codes, namely ultra-sparse regular LDPC codes, which
have a constant variable node degree of $d_{\tv,i} = d_\tv = 2$ and a constant check node degree $d_\tc$. Their design rate
is therefore $1-2/d_\tc$. Previous works have shown that this ultra-sparse structure facilitates 
the design of graphs with a large girth~\cite{venkiah_design_2008}, which is the length of the smallest cycle in the bipartite graph and 
important for iterative decoding. Although their minimum distance 
scales only logarithmically with the blocklength~\cite[Sec.~IV-E]{poulliat_design_2008},
numerical simulation results show lower error floors than their binary counterparts. 
We assume a probability-domain based decoding.

\subsection{PAS with NB Codes: Rate $(m-1)/m$}
\label{sec:simple PAS}

The \ac{PAS} system model is depicted in Fig.~\ref{fig:pas}. It exploits the symmetry property of the optimal input distribution $P_X$ to 
factorize the random variable $X$ into independent random variables referring
to the amplitude and sign (\emph{amplitude-sign factorization}), i.e., $P_X(x) = P_A(\abs{x})P_S(\sign(x))$. The sign distribution $P_S$
is uniform on $\cS=\{-1,+1\}$, while $P_A$ is non-uniform on the amplitude set $\cA = \{1, 3, \ldots, 2^{m}-1\}$. See
\cite[Sec.~III-IV]{bocherer_bandwidth_2015} for further details.

We first use $R_\tc = (m-1)/m$ codes with a $2^m$-ASK constellation.
In Fig. 1, this corresponds to $\gamma=0$. The \ac{DM} maps $k$ data bits to $n$ amplitudes. 
The \ac{FEC} encoder generates redundancy, which is mapped to the $n$ signs. FEC encoding is 
systematic, to preserve the amplitude distribution imposed by the DM. The combination of an amplitude
and a sign results in one channel input symbol. The $n$ channel input symbols can be represented 
by $mn$ bits, which requires an NB code with blocklength $n_\tc=(nm)/p$. 

Each amplitude requires $(m-1)$ bits for its
representation and we require $p = \ell (m-1)$ when $2^m$-ASK \ac{PAS}  
is combined with NB codes over $\setF_{q}$. The variable $\ell\in\setN$ defines the 
number of amplitudes in $\cA$ which are mapped to one $\F_{q}$ symbol. We denote this mapping as 
\begin{equation}
 \beta_{\tA} : \cA^\ell \to \setF_{q}\label{eq:betaA}.
\end{equation}
The amplitude part has a size of $k_\tc = n/\ell$ symbols and is collected in
vector $\vu\in\setF^{k_\tc}_{q}$. Systematic encoding with $\vG = \vect{\vI & \vP}$ yields the parity part $\vp = \vu\vP$ of
$(1-c)n_\tc$ symbols that are approximately uniformly distributed~\cite[Theorem~I]{boutros_probabilistic_2017}. 
We will therefore assume at the decoder that the signs are uniformly distributed. Using the inverse of the mapping
\begin{equation}
 \beta_{\tS} : \{0, 1\}^p \to \setF_{q}\label{eq:betaS}
\end{equation}
we relate each parity symbol to a sign sequence.

For the decoder input we need to calculate the vectors
\begin{equation}
 \vP_i = \vect{P_{C_i|\vY}(0|\vy)\\P_{C_i|\vY}(1|\vy)\\\vdots\\P_{C_i|\vY}\left(\alpha^{q-2}|\vy\right)}, \quad i = 1, \ldots, n_\tc\label{eq:dec_pi}
\end{equation}
where $\alpha$ refers to a primitive element of $\F_q$, i.e., $\F_q = \{0, 1, \alpha, \ldots, \alpha^{q-2}\}$.
The value $P_{C_i|\vY}(c|\vy)$ denotes the probability that the $i$-th codeword symbol is $c$, when $\vy$ was received.

We need to distinguish two cases for the soft-input vectors $\vP_i$ depending on 
whether the codeword symbol $c_i \in \F_{q}$ refers to an amplitude~\eqref{eq:betaA} or sign mapping~\eqref{eq:betaS}. 
Let $\vy^A_i = (y_1, \ldots, y_\ell)$ be the vector of all 
received symbols that resulted from the transmission of the amplitudes associated with the $i$-th codeword symbol. 
Similarly, the vector $\vy^{S}_i = (y_1, \ldots, y_p)$ refers to the received symbols that resulted from 
the transmission of the signs associated with the $i$-th codeword symbol.

\paragraph{Amplitude Mappings}
For $i = 1,\ldots,k_\tc$ and $\va = (a_1, \ldots, a_{\ell}) =  \beta_\tA^{-1}(c)$, assuming uniform signs, the demapper calculates the metric
\begin{align}
 P_{C_i|\vY}(c|\vy^{A}_i) &\propto P_{C_i,\vY}(c,\vy^{A}_i) = P_{\vA\vY}(\beta_\tA^{-1}(c),\vy^{A}_i)\nonumber\\
		    &= \prod_{j=1}^\ell P_{AY}(a_j,y_j)\nonumber\\
                    &= \prod_{j=1}^\ell \sum_{s\in\{\pm 1\}} P_{XY}(a_js,y_j)\nonumber\\
                    &= \prod_{j=1}^\ell \frac{1}{2}P_A(a_j)\sum_{s\in\{\pm 1\}}  p_{Y|X}(y_j|a_js).\label{eq:vnode_input_amplitude}
\end{align}

\paragraph{Sign Mappings}
For the parity part $i = k_\tc+1,\ldots,n_\tc$ and $\vs = \vect{s_1, \ldots, s_p} = \beta_\tS^{-1}(c)$, assuming uniform signs, the
demapper calculates the metric
\begin{align}
  P_{C_i|\vY}(c|\vy^{S}_i) &\propto P_{C_i,\vY}(c,\vy^{S}_i) = P_{\vS\vY}(\beta_\tS^{-1}(c),\vy^{S}_i)\nonumber\\
                      &= \prod_{j=1}^p P_{S_jY}(s_j,y_j)\nonumber\\
                      &= \prod_{j=1}^p \sum_{\substack{x\in\cX:\\ \sign(x) = s_j}} P_{XY}(x,y_j)\nonumber\\
                      &= \prod_{j=1}^p \frac{1}{2}\sum_{a\in\cA} P_{Y|X}(y_j|as_j)P_A(a).\label{eq:vnode_input_sign}
\end{align}

\begin{figure}[h]
 \centering
 \footnotesize
 \includegraphics{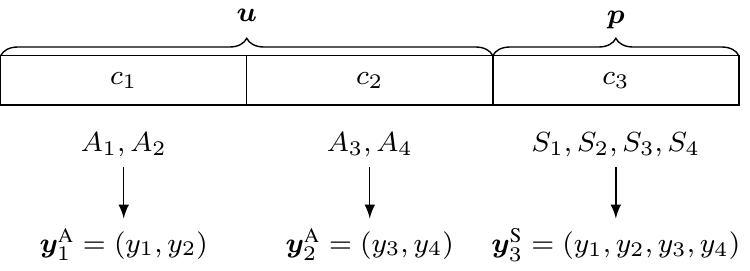}
 \caption{Illustration how the codeword symbols of a $\F_{16}$ code are associated with amplitudes and signs for 8-ASK with PAS. For the codeword
 length of 3 symbols, four channel inputs $X_i=A_iS_i, i=1,\ldots,4$ can be generated.}
 \label{fig:map1}
\end{figure}

We illustrate the setting for $(m-1)/m$ codes and a $2^m$-ASK constellation in
Fig.~\ref{fig:map1} for $m=3$. We consider a code over $\F_{16}$ ($p = 4$) 
and a blocklength of three symbols. Then, each of the two
symbols in the information part represents $\ell = 4/(3-1) = 2$ amplitudes.
The last codeword symbol forms the parity part and is mapped to four sign bits.

\subsection{PAS with NB Codes: Rates Larger Than $(m-1)/m$}
\label{sec:extended_pas}

As for the binary case, \ac{PAS} can also be operated with non-binary codes of rates larger than $(m-1)/m$. In this case, 
$(\gamma n)/p$ information symbols are used as signs, where~\cite[Sec.~IV-D]{bocherer_bandwidth_2015}
\begin{equation}
 \gamma = 1 - (1-c)m\label{eq:def_gamma}.
\end{equation}
This means \eqref{eq:vnode_input_amplitude} must only be applied for the first
$n/\ell$ variables nodes (nodes associated with amplitude mappings) and the remaining $n/p$ variable nodes (nodes associated with 
sign mappings) are initialized with \eqref{eq:vnode_input_sign}.

The overall transmission rate of a \ac{PAS} transmitter is therefore
\begin{equation}
 R_{\tt} = R_\tdm + \gamma\label{eq:transmission_rate}
\end{equation}
and the large flexibility in supported \acp{SE} is achieved by using different \ac{DM} rates $R_\tdm$ for the same \ac{FEC}.

\section{Finite Length Simulations}
\label{sec:numerical_results}

We now present simulation results for ultra-sparse NB-LDPC
codes over $\F_{64}$ and $\F_{256}$. The considered codes are short,
so we must account for the rate loss of the \ac{DM}~\cite{schulte_constant_2016}, which is
\begin{equation}
 R_\tloss = \entr(P_A) - k/n.\label{eq:rate_loss}
\end{equation}
To obtain a desired rate $R_\tt$ and to mitigate the rate loss, we tune the parameter $\nu$ of
the \ac{MB} distribution $P_A(a) \propto \exp(-a^2\nu), a\in\{1,3,\ldots,M-1\}$ to
support the desired rate $R_\tdm = R_\tt - \gamma$.

\subsection{Finite Length Bounds}
\label{sec:finite_length_bounds}

To benchmark the finite length performance we use
Shannon's \ac{SPB}~\cite{shannon_probability_1959} on the average \ac{FER}
$P_\tB$ and Gallager's random coding 
bound~\cite[Theorem~5.6.2]{gallager1968}. The latter is
\begin{equation}
 \E{P_\tB} \leq 2^{-nE_\tG(R_\tt,P_X)}
\end{equation}
where the Gallager exponent is calculated as
{\small
\begin{multline*}
 E_\tG(R_\tt,P_X) =\\ \max_{\rho\in\left[0,1\right]} -\log_2\left(\int\limits_{-\infty}^\infty \left(\sum_{x\in\cX} p_{Y|X}(y|x)^\frac{1}{1+\rho}P_X(x)\right)^{(1+\rho)}\dif{y}\right)-\rho R_\tt.
\end{multline*}
} The distribution $P_X$ is the one chosen to guarantee the desired \ac{SE} $R_\tt$.

\begin{table*}
 \centering
 \footnotesize
 \caption{Summary of the shaped modes used in the numerical simulation results.}
 \label{tab:modes_summary}
 \begin{tabular}{lllllllll}
  \toprule
  Mode & M & $R_\tc$ & $k/n$ & $R_\tdm$ & $\gamma$ & $\SNR_\text{CAP}$ [\si{dB}] & $\SNR_\tsmd$ [\si{dB}] & $\SNR_\tbmd$ [\si{dB}]\\
  \midrule
  8-ASK, $R_{\tt} = \SI{1.5}{\bpcu}$ & 8 & 3/4 & 240/192 & \num{1.25} & 1/4 & \num{8.451} & \num{8.462} & \num{8.484} \\
  8-ASK, $R_{\tt} = \SI{2.0}{\bpcu}$ & 8 & 3/4 & 448/256 & \num{1.75} & 1/4 & \num{11.761} & \num{11.898} & \num{11.920} \\
  16-ASK, $R_{\tt} = \SI{2.75}{\bpcu}$ & 16 & 5/6 & 609/252 & \num{2.4167} & 1/3 & \num{16.460} & \num{16.497} & \num{16.512} \\
  \bottomrule
 \end{tabular}
\end{table*}

\subsection{Numerical Results}

We compare the \ac{FER} of our NB-LDPC codes with the binary LDPC codes suggested for 5G 
by Qualcomm~\cite{qualcomm_codes}. The latter codes are protograph-based~\cite{thorpe_protograph}
and constructed via liftings from a set of three base matrices for low, medium and high code rates. These base matrices have two punctured (state) and degree one variable nodes.
This construction yields a significant performance improvement in the waterfall region~\cite{thorpe_protograph}. Two
hundred \ac{BP} iterations were used. The codes were derived from the high-rate basematrix of the proposal~\cite{qualcomm_codes} and have 
girth 4 ($n_{\tc,\tbin} = \num{576}, R_\tc = 3/4$) and 6 ($n_{\tc,\tbin} = \num{768}, R_\tc = 3/4$; $n_{\tc,\tbin} = \num{1008}, R_\tc = 5/6$).

The non-binary LDPC codes were constructed from protographs of the form
\[
 [\underbrace{2\quad2\quad\ldots\quad 2}_{d_\tc/2}]
\]
via cyclic liftings and a \ac{PEG}-like algorithm~\cite{hu_regular_2005}. All constructed
matrices have girth 8. The coefficients were optimized row-wise by following the binary image approach of~\cite{poulliat_design_2008}.
As in the binary case, we performed a maximum of 200 \ac{BP} iterations for decoding.
The parameters are summarized in Table~\ref{tab:modes_summary}.

\begin{figure}[t]
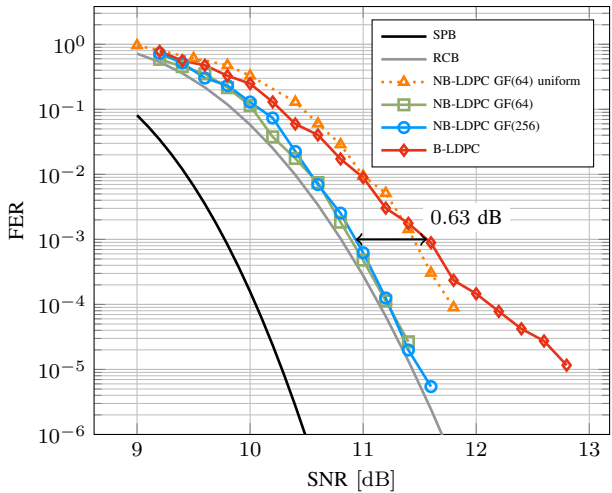

 \centering
 \footnotesize
 \includegraphics{{{figures/coded_results-n=576-SE=1.50}}}
 \caption{Simulation results of suggested NB-LDPC coding scheme for 8-ASK, $R_\tt = \SI{1.5}{\bpcu}$ and binary blocklength of $n_{\tc,\tbin} = \num{576}$ bits.}
 \label{fig:coded_results1}
\end{figure}

Fig.~\ref{fig:coded_results1} shows simulation results for a blocklength of $n_{\tc,\tbin} = \num{576}$ bits and $R_\tt = \SI{1.5}{\bpcu}$. We first compare
two $\F_{64}$ codes ($n_\tc = \num{96}$ symbols), where the green line with squares refers to the shaped scenario with a rate $3/4$ ($d_\tc = 8$) code, whereas 
the orange line with triangles is its uniform counterpart of rate 1/2 ($d_\tc = 4$). The shaped case clearly improves  over the uniform one.
We also show a curve for a $\F_{256}$ code of rate $3/4$ which has almost the same performance as the $\F_{64}$ version. These results are complemented by
a Monte Carlo \ac{DE} analysis which yields asymptotic decoding thresholds of \SI{9.54}{dB} (uniform), \SI{8.76}{dB} (shaped, $\F_{64}$) and \SI{8.79}{dB}
(shaped, $\F_{256}$). The latter two thresholds only exhibit a gap of about \SI{0.3}{dB} to the respective Shannon limit (cf. Table~\ref{tab:modes_summary}).
Comparing both NB approaches to the binary LDPC code, we see an improvement of \SI{0.63}{dB} at a \ac{FER} of \num{e-3}. The same \ac{DM} with
the same shaping parameters is used for both settings. We emphasize that the gain is not due to the different decoding metrics (\ac{BMD} vs. \ac{SMD}) 
as suggested by the asymptotic Shannon limits in Table~\ref{tab:modes_summary}.

\begin{figure}[t]
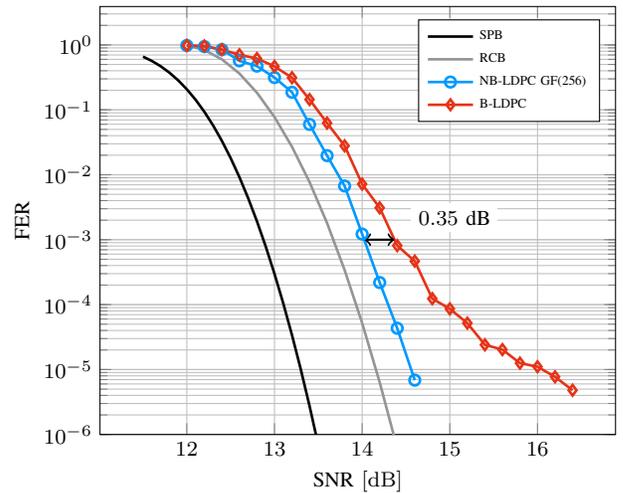

 \centering
 \footnotesize
 \includegraphics{{{figures/coded_results-n=768-SE=2.00}}}
 \caption{Simulation results of suggested NB-LDPC coding scheme for 8-ASK, $R_\tt = \SI{2.0}{\bpcu}$ and binary blocklength of $n_{\tc,\tbin} = \num{768}$ bits.}
 \label{fig:coded_results2}
\end{figure}

Fig.~\ref{fig:coded_results2} shows simulation results for $n_{\tc,\tbin} = \num{768}$ bits and $R_\tt = \SI{2.0}{\bpcu}$. As before, we see a clear improvement
of \SI{0.35}{dB} over the binary LDPC code. 

\begin{figure}[t]
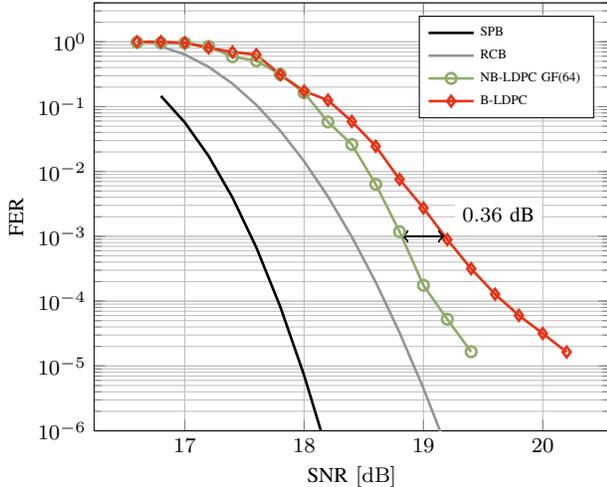

 \centering
 \footnotesize
 \includegraphics{{{figures/coded_results-n=1008-SE=2.75}}}
 \caption{Simulation results of suggested NB-LDPC coding scheme for 16-ASK, $R_\tt = \SI{2.75}{\bpcu}$ and binary blocklength of $n_{\tc,\tbin} = \num{1008}$ bits.}
 \label{fig:coded_results3}
\end{figure}

Finally, Fig.~\ref{fig:coded_results3} depicts the performance for $n_{\tc,\tbin} = 1008$ bits and $R_\tt = \SI{2.75}{\bpcu}$. Here, we use a 16-ASK constellation and
code rate $5/6$. The observations from Figs.~\ref{fig:coded_results1} and \ref{fig:coded_results2} also carry on to higher \acp{SE}.

\section{Conclusion}
\label{sec:conclusion}

In this paper, we showed how to combine non-binary codes over $\F_{2^p}$ with \ac{PAS}. Numerical
simulation results with ultra-sparse high order \ac{NB-LDPC} codes show a significant improvement of up to \SI{0.63}{dB} over state-of-the-art binary \ac{LDPC} codes
combined with PAS for the shortest considered blocklength of \num{576} bits at a \ac{FER} of \num{e-3}. For future research, we plan to tackle the issue of code design 
using a protograph based approach to take the different variable node input
\acp{PMF} into account (see~\eqref{eq:vnode_input_amplitude}, \eqref{eq:vnode_input_sign}). For instance, we combine the approaches of \cite{bennatan_design_2006, 
chang_exit_2011,steiner_protograph-based_2016}. Additional work should also focus on code designs for smaller field sizes
to decrease the decoding complexity, see e.g., \cite{dolecek_non-binary_2014}.


\end{document}